\documentclass[letter]{aa} 
\usepackage{graphicx}
\usepackage{txfonts}
\begin{document} 

   \title{L'-band AGPM vector vortex coronagraph's first light on VLT/NACO}
   \subtitle{Discovery of a late-type companion at two beamwidths from an F0V star}

  \author{D. Mawet\inst{1}
\and O. Absil\inst{2}
\and C. Delacroix\inst{2} 
\and J. H. Girard\inst{1}
\and J. Milli\inst{1}
\and J. O'Neal\inst{1}
\and P. Baudoz\inst{5}
\and A. Boccaletti\inst{5}
\and P. Bourget\inst{1}
\and V. Christiaens\inst{2}
\and P. Forsberg\inst{4}
\and F. Gonte\inst{1}
\and S. Habraken\inst{2}
\and C. Hanot\inst{2}
\and M. Karlsson\inst{4}
\and M. Kasper\inst{3}
\and J.-L. Lizon\inst{3}
\and K. Muzic\inst{1}
\and R. Olivier\inst{6}
\and E. Pe\~na\inst{1}
\and N. Slusarenko\inst{1}
\and L.~E. Tacconi-Garman\inst{3}
\and J. Surdej\inst{2}
}

\institute{European Southern Observatory Vitacura, Alonso de Cordova 3107, Casilla 19001, Vitacura, Santiago 19, Chile
\and D\'epartement d'Astrophysique, G\'eophysique et Oc\'eanographie, Universit\'e de Li\`ege, 17 All\'ee du Six Ao\^ut, B-4000 Li\`ege, Belgium
\and European Southern Observatory Headquarters, Karl-Schwarzschild-Str. 2, 85748 Garching bei M\"{u}nchen, Germany
 \and {Department of Engineering Sciences, \AA}ngstr\"{o}m Laboratory, Uppsala University, Box 534, SE-751 21 Uppsala, Sweden
 \and LESIA, Observatoire de Paris, 5 pl. J. Janssen, F-92195 Meudon, France
\and GDTech s.a., LIEGE Science Park, rue des Chasseurs Ardennais, B-4031 Li\`ege, Belgium
}

   \date{Accepted April 1, 2013}

 
  \abstract
   {High contrast imaging has thoroughly combed through the limited search space accessible with first-generation ground-based adaptive optics instruments and the Hubble Space Telescope. Only a few objects were discovered, and many non-detections reported and statistically interpreted. The field is now in need of a technological breakthrough.}
   {Our aim is to open a new search space with first-generation systems such as NACO at the Very Large Telescope, by providing ground-breaking inner working angle (IWA) capabilities in the L' band. The L' band is a sweet spot for high contrast coronagraphy since the planet-to-star brightness ratio is favorable, while the Strehl ratio is naturally higher.}
   {An annular groove phase mask (AGPM) vector vortex coronagraph optimized for the L' band, made from diamond subwavelength gratings was manufactured and qualified in the lab. The AGPM enables high contrast imaging at very small IWA, potentially being the key to unexplored discovery space.}
   {Here we present the installation and successful on-sky tests of an L'-band AGPM coronagraph on NACO. Using angular differential imaging, which is well suited to the rotational symmetry of the AGPM, we demonstrated a $\Delta L' > 7.5$ mag~contrast from an IWA $\simeq 0\farcs 09$ onwards, during average seeing conditions, and for total integration times of a few hundred seconds. }
   {}
   \maketitle
%

\section{Introduction}
The goal of high contrast imaging is primarily to discover and characterize extra-solar planetary systems. For technical motives rather than scientific ones, most surveys have targeted young and nearby stars. This search space is already limited, and imaging surveys have only explored its surface, strongly limited by contrast and inner working angle (IWA) capabilities \citep{AbsilMawet2010}. Despite the few spectacular objects discovered and insightful lessons learned from the majority of non detections, hunting for long-period planets has mostly been a hard and unfruitful task, with a very low yield \citep{Lafreniere2007b,Chauvin2010,Vigan2012}. It is expected that opening the parameter space to fainter/smaller planets closer to their parent stars will bring many new objects \citep{Crepp2011}. The new sample is critical to exoplanet science because it will shed some light on planet formation mechanisms at or within the snow line, and help bridge the gap between the population of close planets discovered by radial velocity or transit techniques and the free-floating planets discovered by microlensing observations \citep{Quanz2012}. This intermediate parameter space should be opened by the second-generation coronagraphic instruments that have started to arrive at major observatories \citep{Macintosh2012,Kasper2012,Oppenheimer2012,Martinache2012}. Coronagraphy promises to be high contrast imaging's sharpest tool, but requires exquisite image quality and stability to perform efficiently. These new instruments have thus been designed accordingly. However, first-generation instruments still possess untapped potential that only ten years of operations and understanding allow us to exploit fully \citep{Girard2012}, especially in the mid-infrared (L' band, from 3.5 to 4.2 $\mu$m). This wavelength range offers significant advantages compared to shorter wavelengths \citep{Kasper2007}: (i) The L'-band contrast of planetary-mass companions with respect to their host stars is predicted to be more favorable than in the H band \citep{Baraffe2003,Fortney2008, Spiegel2012} so that lower-mass, older objects can be addressed; and (ii) the L'-band provides better and more stable image quality, with Strehl ratios well above 70\% and sometimes as high as 90\%, thus reducing speckle noise. These advantages certainly compensate for the increased sky background in the thermal infrared and the loss in resolution, especially if small IWA phase-mask coronagraphs are available. 

Here we describe the successful implementation of an L'-band annular groove phase mask \citep[AGPM;][]{Mawet2005b} vector vortex coronagraph on NACO \citep{Lenzen2003,Rousset2003}, the adaptive optics instrument of ESO's Very Large Telescope (VLT). To our knowledge it is the first time that an image-plane phase-mask coronagraph has been used in the mid-infrared.  

\begin{figure}[t]
  \centering
\includegraphics[scale=0.85]{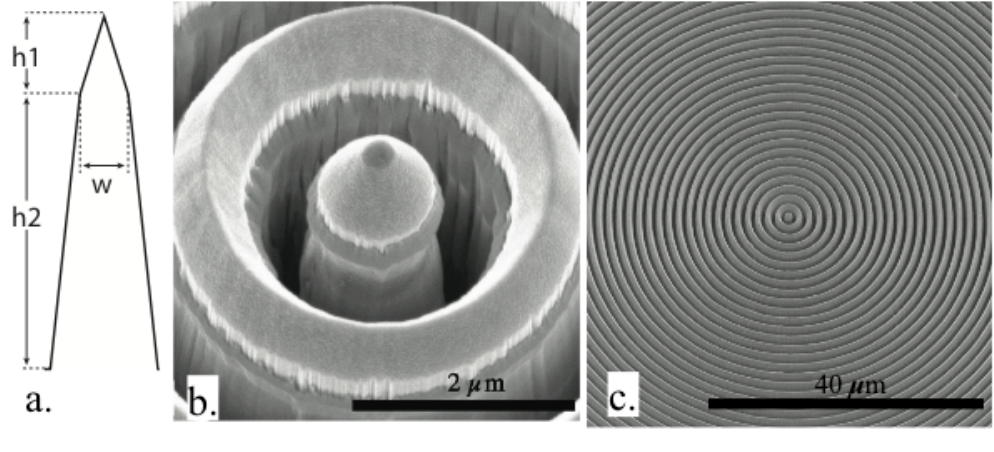}
  \caption{Scanning electron microscope (SEM) images of the NACO AGPM. a. Structure profile schematic, with h$_2=5\pm 0.1$ $\mu$m, h$_1=1\pm 0.1$ $\mu$m, and w $=0.65\pm 0.03$ $\mu$m (the grating pitch is 1.42 $\mu$m). b. Zoom on the center of the diamond AGPM. c. Overview of the structure showing the uniformity and original cleanliness of this particular device. \label{fig0}}
\end{figure}

\section{An AGPM vector vortex coronagraph on NACO}
The AGPM is an optical vortex made from diamond subwavelength gratings (Fig.~\ref{fig0}). When centered on the diffraction pattern of a star seen by a telescope, optical vortices affect the subsequent propagation to the downstream Lyot stop by redirecting the on-axis starlight outside the pupil \citep[e.g.,][]{Mawet2005b}. The advantages of the AGPM coronagraph over classical Lyot coronagraphs or phase/amplitude apodizers are small IWA, down to $0.9\lambda/D$ (e.g., $0\farcs 09$ in the L' band at the VLT, slightly smaller than the diffraction limit); clear 360$^\circ$ off-axis field of view/discovery space; outer working angle set only by the instrument and/or mechanical/optical constraints; achromatic over the entire working waveband (here L' band); high throughput (here $\simeq 88\%$); and optical/operational simplicity. After eight years of intense technological development, the AGPM has reached a sufficient readiness level for telescope implementation \citep{Delacroix2013,Forsberg2013a}. The AGPM selected for NACO was the third one in a series of four realizations (AGPM-L3). Its theoretical raw null depth limited by its intrinsic chromatism was estimated (assuming a trapezoidal profile,\footnote{This AGPM differs from the one tested in \citet{Delacroix2013}. The tops of the grating walls are triangular to improve the transmittance (Fig.~\ref{fig0}). This profile was etched with a process similar to the fabrication of broadband antireflective structures \citep{Forsberg2013b}.} see Fig.~\ref{fig0}) and measured to be around $5\times 10^{-3}$ (corresponding to a raw contrast of $2.5 \times 10^{-5}$ at $2\lambda/D$), which is more than needed for on-sky operations where the limit is set by the residual wavefront aberrations.
\begin{figure}[t]
  \centering
\includegraphics[scale=0.31]{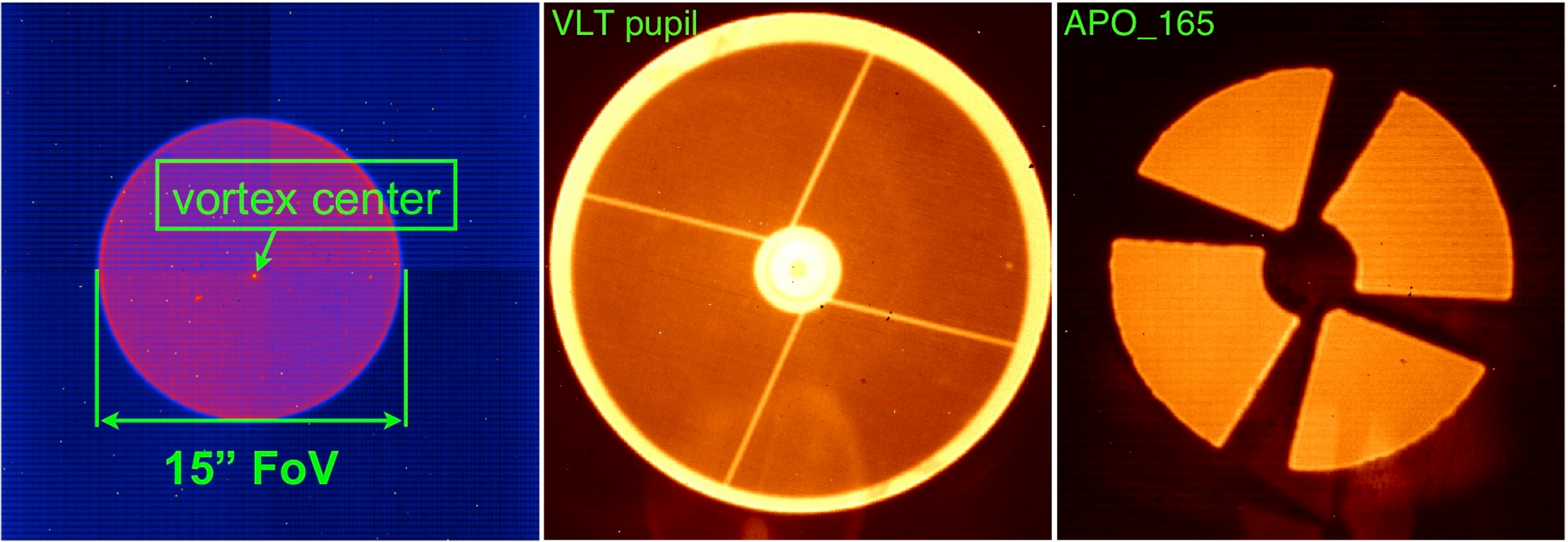}
  \caption{Left: view from inside CONICA, showing the 9 mm clear aperture corresponding to a 15\arcsec field of view (diameter), fully contained within the 27\arcsec field of view of the L27 objective. Middle: full oversized stop of CONICA, showing the VLT pupil (including the central obscuration and struts). Right: APO165 pupil mask (diameter$=0.87\times D_{pup}$) available inside CONICA, aligned to cover the diffraction and thermal background from the central obscuration and struts. \label{fig1}}
\end{figure}

The AGPM was installed inside NACO as part of a planned overhaul in November 2012. The AGPM was mounted on the entrance slit wheel by means of a dedicated aluminum mount, designed by GDTech s.a. The assembly of the mount and AGPM was done on site at Paranal observatory in a clean room environment. Prior to on-sky tests and operations, a CONICA internal image of the mask was done (see Fig.~\ref{fig1}, left), revealing significant dust contamination, marginally affecting the background noise. The slit wheel was set so that the center of the AGPM falls close to but slightly away from CONICA's detector quadrant intersection. The AGPM field of view is $\simeq 15\arcsec$, corresponding to an outer working angle (OWA) of $7\farcs 5$. The OWA is only limited by the size of the device (10 mm in diameter) and its mount. The mask transmittance at L' band was measured on the sky to be $85\% \pm 5\%$, which is consistent with the theoretical value and laboratory measurements, both $\simeq 88\%$, limited by imperfect antireflective treatments and mild absorption features around 4 $\mu$m \citep{Delacroix2013}.
To stabilize speckles, we used the pupil tracking mode enabling angular differential imaging \citep[ADI,][]{Marois2006}, which is perfectly adapted to the circular symmetry and 360$^\circ$ field of view of the AGPM. The CONICA camera is equipped with a pupil mask which blocks the telescope central obscuration and spiders. Once correctly aligned with the pupil (in x, y, and $\theta$), this mask is optimal for use with the AGPM in pupil tracking mode (see Fig.~\ref{fig1}). The measured throughput of the APO165 mask used here is $\simeq 60\%$. In terms of sensitivity, it is worth noting that the throughput loss is almost entirely compensated by the improved thermal background. The pupil obscuration is responsible for more than 25\% of the thermal emissivity of the telescope, even though its area only covers $\simeq 5\%$. Therefore, instead of loosing $1-\sqrt{0.6}\approx 0.225$ in sensitivity, only $1-0.6/\sqrt{0.6*0.75}\approx 0.1$ is actually lost. To maintain its high contrast capabilities, the centering of the star on the AGPM must be within $(\lambda/D)/10$, or $\simeq 10$ mas (a third of a pixel at L27). This level of centering capability is now routinely obtained with NACO, both in pupil and field tracking modes; we typically measure $\simeq 1$ mas/minute drifts across the meridian.
 \begin{table}[t]
      \caption[]{Observing log.}
         \label{table0}
	$$
         \begin{array}{p{0.25\linewidth}ll}
            \hline \hline
            \noalign{\smallskip}
         Date				& 09/12/2012  				& 11/02/2013										\\
         Star				& {\rm HD4691}			& {\rm HD123888} 									\\
	Spectral type		& {\rm F0V}				& {\rm K1III}										\\
	V mag			& 6.79					& 6.62          										\\
	L app.~mag		& 5.86					& 4.01 											\\ 
     \noalign{\smallskip}
            \hline
            \noalign{\smallskip}      
          DIT/NDIT/\# fr		& 0.2{\rm s}/10/100 			& 0.25{\rm s}/80/40									\\
          Seeing            		& 1\arcsec-1\farcs 5  				& 0\farcs 8-0\farcs 9    			 \\
          $\tau_0$ 			& 2-4 {\rm ms}  				& 4-5 {\rm ms}    			 \\
          Strehl ratio		&65-80\% 				& 75-80\% 										\\  
         PA range			&\simeq 30^\circ 			&\simeq 30^\circ 									\\
            \noalign{\smallskip}
            \hline
         \end{array}
	$$
   \end{table}

\begin{figure*}[t]
  \centering
\includegraphics[scale=0.65]{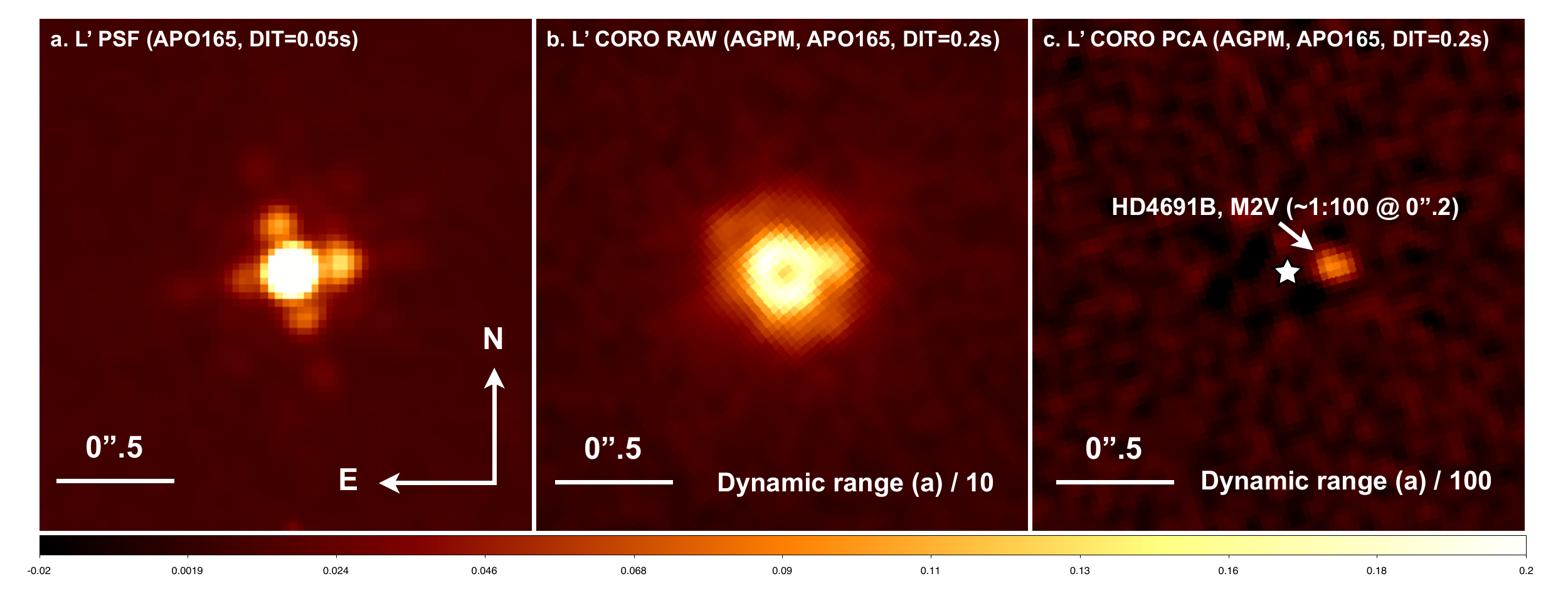}
  \caption{a: L'-band NACO PSF with the APO165 pupil mask in the beam. b: L'-band NACO coronagraphic image with the star centered on the AGPM (the dynamic range and corresponding colorbar/scale are a factor of 10 smaller than in image a). c: Result of our PCA-ADI data reduction pipeline, revealing a putative faint off-axis M2V companion at only $\simeq 0\farcs 19$ (the dynamic range and corresponding colorbar/scale are a factor of 100 smaller than in image a). The scale is linear on all images (and scaled down by a factor of 10 and 100 for b and c, respectively), illustrating in sequence the benefits of coronagraphy and optimized data reduction technique. The bottom color bar refers to Figure a. \label{fig3}}
\end{figure*}

\section{First light}
On December 9, 2012, a representative observing sequence was performed on the 1.9-Gyr old main sequence standard star HD4691 (see Table~\ref{table0}), under $\simeq 1\farcs 2$ visual seeing conditions. This star was chosen to maximize brightness and field rotation during the short time allocated for this technical test. A $\simeq 30$-minute ADI sequence was obtained with a parallactic angle (PA) range of 30$^\circ$ and for a total exposure time of 200s on source; the efficiency was mediocre for technical reasons. After acquiring an off-axis PSF for photometric reference, we measured an instantaneous contrast of $\simeq 50$ peak-to-peak (despite the average-to-bad conditions, see Table~\ref{table0}). The attenuation is about 5 times higher than measured with NACO's four-quadrant phase-mask coronagraph at Ks \citep{Boccaletti2004}. The coronagraph diffraction control yields two instantaneous benefits compared to classical imaging: (i) the peak saturation limit is decreased by a factor $\simeq 50$, and (ii) the level of quasi-static speckles pinned to the PSF and the stellar photon noise limit are potentially decreased by a factor $\simeq \sqrt{50} \approx 7$, both within the AO control radius of $7\lambda/D$. All in all, the L'-band AGPM coronagraph allows the background limit to be reached much closer in.

\subsection{Detection of a candidate companion}
After applying basic cosmetic treatment to our sequence of 100 frames (background subtraction, flat fielding, and bad pixel/cosmic ray correction), we decided to use the quality and stability of the L'-band PSF provided by NACO to perform a sophisticated speckle subtraction. We used the very efficient principal component analysis (PCA) algorithm presented in \citet{Soummer2012}. The result, using the whole image and retaining three main components, is presented in Fig.~\ref{fig3}. By pure chance, the object has a $\simeq 1:100$ (or $\Delta L' \simeq 5$) off-axis companion located at $\simeq 0\farcs 19$ ($< 2 \lambda/D$), making this our first unexpected scientific result. The companion flux and astrometry were obtained by using the fake negative companion technique \citep{Marois2010}. The method proceeds as follows: (i) estimate the (biased) position and flux of the companion from the first reduced image; (ii) use the measured off-axis PSF as a template to remove this first estimate from the cleaned data cube before applying PCA; and (iii) iterate on the position x,y and flux until a well-chosen figure of merit is minimized ($\chi^2$ in a pie chart aperture centered on the first estimate of the companion position, 2.44$\lambda/D$ in radius and $6\times 1.22\lambda/D$ in azimuth). The minimization was performed with the Simplex-Amoeba optimization. 

Close to the center where the speckle field is intense, the companion flux can be overestimated because the minimization tries to subtract underlying speckles. To estimate our error bars, we decided to proceed with an alternative method called smart-ADI PCA: the frames used to construct the component basis are selected according to a minimum azimuthal separation criterion (here $N_\delta = 0.7 \lambda/D$). With this technique, flux is much better preserved. We measured up to 200\% additional flux compared to normal ADI PCA. However, fake planet tests still indicate a 25\% flux loss, confirming that this method underestimates the flux and thus provides our lower bound. Finally, the coronagraph off-axis attenuation profile, measured in the lab \citep{Delacroix2013} was also taken into account. Using the BCAH98 model \citep{Baraffe1998}, we derived the properties of the newly discovered candidate companion, assuming association. Note that the TRILEGAL starcount model \citep{Girardi2005} yields a probability of $4\times 10^{-7}$ that it is an unrelated background object. At an absolute L'-band magnitude of 6.65, and 1.9 Gyr for the system, the putative companion would most likely be an M2V star at projected separation of $11.8 \pm 0.4$ AU, and $354\fdg 5 \pm 0\fdg 6$ position angle (see Table \ref{table1}). 

\subsection{Representative NACO AGPM detection limits}
\begin{figure}[t]
  \centering
\includegraphics[scale=0.4]{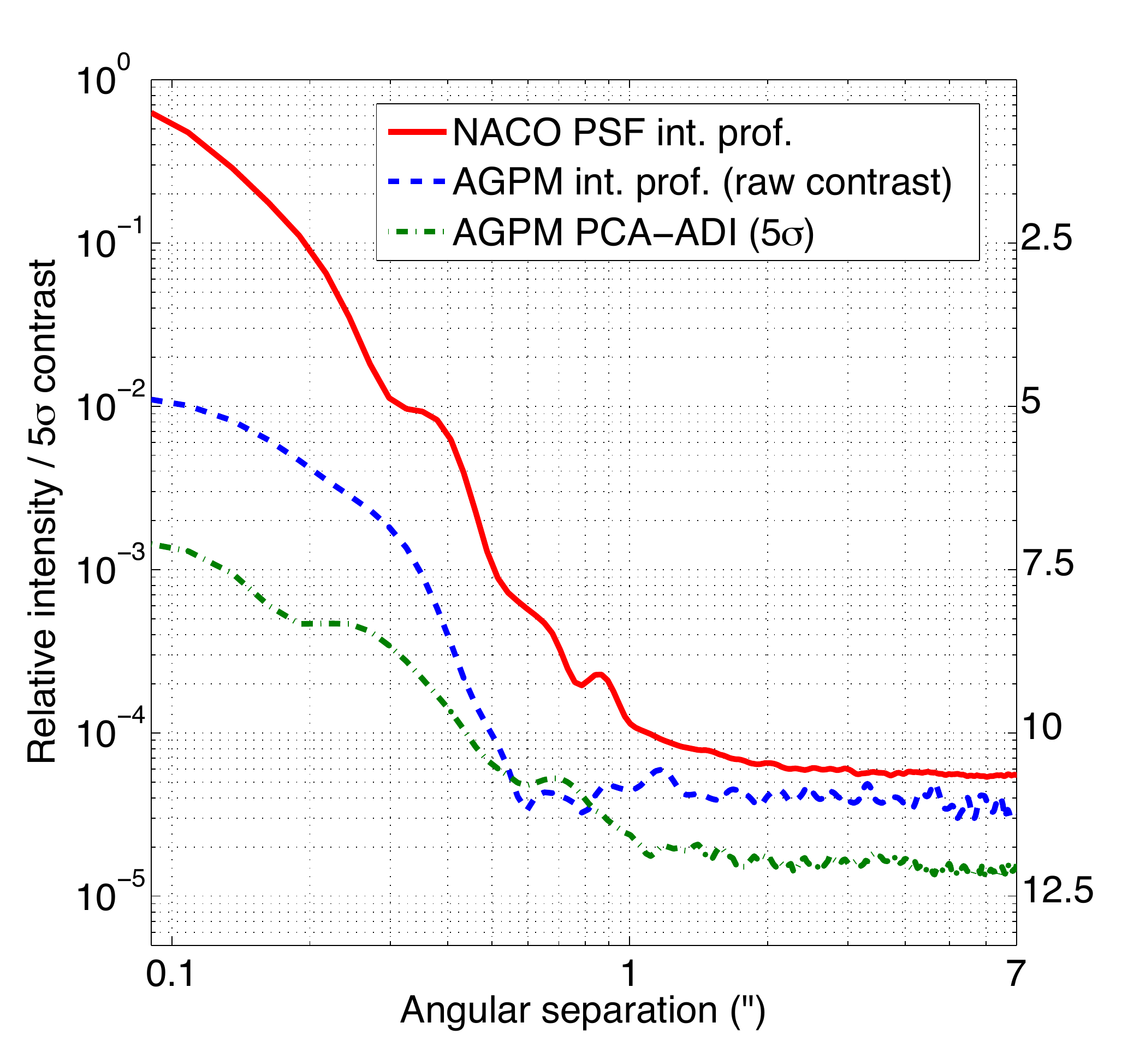}
  \caption{Normalized azimuthally averaged relative intensity profiles and contrast curve. The plain red curve shows the intensity profile of a typical saturated NACO L' PSF (similar brightness and exposure time). The blue dashed curve shows the AGPM intensity profile before PCA, demonstrating the instantaneous contrast gain provided by the coronagraph at all spatial frequencies within the AO control radius ($\simeq 0\farcs 7$). The green dash-dot curve presents the reduced PCA-ADI $5\sigma$ detectability limits (40 frames, 800s, $\Delta PA\approx 30^\circ$), taking both the coronagraph off-axis transmission and the PCA-ADI flux losses into account.  \label{fig4}}
\end{figure}

Since the presence of the companion affects the contrast, we took another similar representative ADI sequence on a different standard star (HD123888, see Table~\ref{table0}). This technical test was performed under better conditions, and benefited from our improved mastering of the new mode (efficiency was four times better than during the first light). A similar instantaneous attenuation was confirmed. To calibrate our detection limits against flux losses induced by PCA, we injected fake companions (at $15\sigma$) prior to PCA and measured their throughput after PCA. We used the derived throughput map to renormalize the initial contrast curve a posteriori (see Fig.~\ref{fig4}). Figure~\ref{fig4} shows excellent detection capabilities down to the IWA of the AGPM. The final calibrated contrast presented here (green dash-dot curve), is limited by the small PA range, especially at small angles. The floor reached beyond $1\arcsec$ is due to the background at L', and will be lower for brighter targets and/or longer integrations.

We would like to raise several flags that we will thoroughly address in subsequent papers: (i) Classical tools assuming Gaussian statistics, perfectly valid at large separation, lose significance close to the center simply because the sample size decreases dramatically. At a given angular separation $r$ (in $\lambda/D$), there are $2\pi r$ resolution elements, i.e., only 6 at $r=1\lambda/D$, 12 at $r=2\lambda/D$, etc. (ii) The probability density function (PDF) of speckle noise and associated confidence level for detection depend on radius. ADI was shown to transform speckles' modified Rician PDF into quasi-Gaussian PDF at large separations, but it  is expected that this property of ADI does not hold true at small angles \citep{Marois2008}. (iii) The flux attenuation induced by ADI, potentially significant at small angles, does not scale linearly with the companion brightness, which makes its calibration more difficult. These points should be kept in mind when interpreting contrast curves such as those presented in Fig.~\ref{fig4}, but also all contrast/detectivity plots that have been presented so far in the literature for very small angles.

\begin{table}
      \caption[]{Properties of HD4691B.}
         \label{table1}
	$$
         \begin{array}{p{0.5\linewidth}l}
            \hline            \hline
            \noalign{\smallskip}
         Distance			& 62 {\rm \, pc} \\ 
	Age	\citep{Holmberg2009}      & 1.9 {\rm \, Gyr} \\ 
	Companion abs.~L'~mag	       & 6.65 \pm 0.1 \\ 
   	Mass				& \simeq 0.3 {\rm \, M}_{\odot} \\
   	Temperature			& \simeq 3450 {\rm \, K} \\
	Sp Type			& {\rm \, M2V} \\
	Angular sep. 	       & 0\farcs 190 \pm 0\farcs 007\\
	Proj.~angular sep. 	       & 11.8 \pm 0.4 {\rm \, AU}	\\
	Position angle			&279\degr \pm 0\fdg 6 \\
            \noalign{\smallskip}
            \hline
         \end{array}
	$$
   \end{table}

\section{Conclusions}
The AGPM was designed to provide exquisite IWA (and OWA) capabilities, down to $0.9\lambda/D$ ($0\farcs 09$ at L'), as demonstrated in Fig.~\ref{fig4}. The downside of the AGPM's small IWA is its sensitivity to the Strehl ratio (as all coronagraphs) and to pointing errors. The Apodizing Phase Plate (APP) is another advanced coronagraph offered at L' \citep{Quanz2010,Kenworthy2013}. The only, but significant, benefit of this pupil plane phase apodizer over the AGPM is its intrinsic immunity to tip-tilt errors. This advantage, which has to be traded off with the significantly limited field of view provided by the APP, is decisive when tip-tilt is an issue as was the case with NACO prior to November 2011 \citep{Girard2012}. However, it is less obvious when the instrument provides nominal PSF stability.

In a single technical run, the L'-band AGPM has proved to be a reliable coronagraphic solution, and one of the best high-contrast imaging modes of NACO (and most likely worldwide). Combined with ADI, we demonstrated that high contrast of the order of $\Delta L' > 7.5$ mag~can be reached from the IWA of $0\farcs 09$ onwards, even with very modest on-source integration time, PA variation, and average conditions. The field of view is a clear 360$^\circ$ discovery space 15$\arcsec$ in diameter. The coronagraph is optimized for pupil tracking and is easy to use, thanks to the stability of the NACO L'-band PSF.

\begin{acknowledgements}
 This work was carried out at the European Southern Observatory (ESO) site of Vitacura (Santiago, Chile). OA and JS acknowledge support from the Communaut\'e fran\c{c}aise de Belgique - Actions de recherche concert\'ees - Acad\'emie universitaire Wallonie-Europe. We would like to thank the referee, Dr Christian Marois, for his constructive comments.

\end{acknowledgements}

\bibliographystyle{aa} 
\bibliography{naco_agpm_onsky_dm_letter_astroph} 

\begin{thebibliography}{31}
\expandafter\ifx\csname natexlab\endcsname\relax\def\natexlab#1{#1}\fi

\bibitem[{{Absil} \& {Mawet}(2010)}]{AbsilMawet2010}
{Absil}, O. \& {Mawet}, D. 2010, \aapr, 18, 317

\bibitem[{{Baraffe} {et~al.}(1998){Baraffe}, {Chabrier}, {Allard}, \&
  {Hauschildt}}]{Baraffe1998}
{Baraffe}, I., {Chabrier}, G., {Allard}, F., \& {Hauschildt}, P.~H. 1998, \aap,
  337, 403

\bibitem[{{Baraffe} {et~al.}(2003){Baraffe}, {Chabrier}, {Barman}, {Allard}, \&
  {Hauschildt}}]{Baraffe2003}
{Baraffe}, I., {Chabrier}, G., {Barman}, T.~S., {Allard}, F., \& {Hauschildt},
  P.~H. 2003, \aap, 402, 701

\bibitem[{{Boccaletti} {et~al.}(2004){Boccaletti}, {Riaud}, {Baudoz},
  {Baudrand}, {Rouan}, {Gratadour}, {Lacombe}, \& {Lagrange}}]{Boccaletti2004}
{Boccaletti}, A., {Riaud}, P., {Baudoz}, P., {et~al.} 2004, \pasp, 116, 1061

\bibitem[{{Chauvin} {et~al.}(2010){Chauvin}, {Lagrange}, {Bonavita},
  {Zuckerman}, {Dumas}, {Bessell}, {Beuzit}, {Bonnefoy}, {Desidera}, {Farihi},
  {Lowrance}, {Mouillet}, \& {Song}}]{Chauvin2010}
{Chauvin}, G., {Lagrange}, A.-M., {Bonavita}, M., {et~al.} 2010, \aap, 509, A52

\bibitem[{{Crepp} \& {Johnson}(2011)}]{Crepp2011}
{Crepp}, J.~R. \& {Johnson}, J.~A. 2011, \apj, 733, 126

\bibitem[{{Delacroix} {et~al.}(2013){Delacroix}, {Absil}, {Forsberg}, {Mawet},
  {Christiaens}, {Karlsson}, \& {Boccaletti}}]{Delacroix2013}
{Delacroix}, C., {Absil}, O., {Forsberg}, P., {et~al.} 2013, accepted to \aap

\bibitem[{{Forsberg} \& {Karlsson}(2013{\natexlab{a}})}]{Forsberg2013a}
{Forsberg}, P. \& {Karlsson}, M. 2013{\natexlab{a}}, Diamond \& Related
  Materials, 34, 19

\bibitem[{{Forsberg} \& {Karlsson}(2013{\natexlab{b}})}]{Forsberg2013b}
{Forsberg}, P. \& {Karlsson}, M. 2013{\natexlab{b}}, Optics Express, 21, 2693

\bibitem[{{Fortney} {et~al.}(2008){Fortney}, {Marley}, {Saumon}, \&
  {Lodders}}]{Fortney2008}
{Fortney}, J.~J., {Marley}, M.~S., {Saumon}, D., \& {Lodders}, K. 2008, \apj,
  683, 1104

\bibitem[{{Girard} {et~al.}(2012){Girard}, {O'Neal}, {Mawet}, {Kasper}, {Zins},
  {Neichel}, {Kolb}, {Christiaens}, \& {Tourneboeuf}}]{Girard2012}
{Girard}, J.~H.~V., {O'Neal}, J., {Mawet}, D., {et~al.} 2012, in Proc. SPIE,
  Vol. 8447

\bibitem[{{Girardi} {et~al.}(2005){Girardi}, {Groenewegen}, {Hatziminaoglou},
  \& {da Costa}}]{Girardi2005}
{Girardi}, L., {Groenewegen}, M.~A.~T., {Hatziminaoglou}, E., \& {da Costa}, L.
  2005, \aap, 436, 895

\bibitem[{{Holmberg} {et~al.}(2009){Holmberg}, {Nordstr{\"o}m}, \&
  {Andersen}}]{Holmberg2009}
{Holmberg}, J., {Nordstr{\"o}m}, B., \& {Andersen}, J. 2009, \aap, 501, 941

\bibitem[{Kasper {et~al.}(2007)Kasper, Apai, Janson, \& Brandner}]{Kasper2007}
Kasper, M., Apai, D., Janson, M., \& Brandner, W. 2007, Astronomy and
  Astrophysics, 472, 321

\bibitem[{{Kasper} {et~al.}(2012){Kasper}, {Beuzit}, {Feldt}, {Dohlen},
  {Mouillet}, {Puget}, {Wildi}, {Abe}, {Baruffolo}, {Baudoz}, {Bazzon},
  {Boccaletti}, {Brast}, {Buey}, {Chesneau}, {Claudi}, {Costille},
  {Delboulb{\'e}}, {Desidera}, {Dominik}, {Dorn}, {Downing}, {Feautrier},
  {Fedrigo}, {Fusco}, {Girard}, {Giro}, {Gluck}, {Gonte}, {Gojak}, {Gratton},
  {Henning}, {Hubin}, {Lagrange}, {Langlois}, {Mignant}, {Lizon}, {Lilley},
  {Madec}, {Magnard}, {Martinez}, {Mawet}, {Mesa}, {M{\"u}ller-Nilsson},
  {Moulin}, {Moutou}, {O'Neal}, {Pavlov}, {Perret}, {Petit}, {Popovic},
  {Pragt}, {Rabou}, {Rochat}, {Roelfsema}, {Salasnich}, {Sauvage}, {Schmid},
  {Schuhler}, {Sevin}, {Siebenmorgen}, {Soenke}, {Stadler}, {Suarez},
  {Turatto}, {Udry}, {Vigan}, \& {Zins}}]{Kasper2012}
{Kasper}, M., {Beuzit}, J.-L., {Feldt}, M., {et~al.} 2012, The Messenger, 149,
  17

\bibitem[{{Kenworthy} {et~al.}(2013){Kenworthy}, {Meshkat}, {Quanz}, {Girard},
  {Meyer}, \& {Kasper}}]{Kenworthy2013}
{Kenworthy}, M.~A., {Meshkat}, T., {Quanz}, S.~P., {et~al.} 2013, \apj, 764, 7

\bibitem[{{Lafreni{\`e}re} {et~al.}(2007){Lafreni{\`e}re}, {Doyon}, {Marois},
  {Nadeau}, {Oppenheimer}, {Roche}, {Rigaut}, {Graham}, {Jayawardhana},
  {Johnstone}, {Kalas}, {Macintosh}, \& {Racine}}]{Lafreniere2007b}
{Lafreni{\`e}re}, D., {Doyon}, R., {Marois}, C., {et~al.} 2007, \apj, 670, 1367

\bibitem[{{Lenzen} {et~al.}(2003){Lenzen}, {Hartung}, {Brandner}, {Finger},
  {Hubin}, {Lacombe}, {Lagrange}, {Lehnert}, {Moorwood}, \&
  {Mouillet}}]{Lenzen2003}
{Lenzen}, R., {Hartung}, M., {Brandner}, W., {et~al.} 2003, in Proc. SPIE, ed.
  M.~{Iye} \& A.~F.~M. {Moorwood}, Vol. 4841, 944--952

\bibitem[{{Macintosh} {et~al.}(2012){Macintosh}, {Anthony}, {Atwood},
  {Barriga}, {Bauman}, {Caputa}, {Chilcote}, {Dillon}, {Doyon}, {Dunn},
  {Gavel}, {Galvez}, {Goodsell}, {Graham}, {Hartung}, {Isaacs}, {Kerley},
  {Konopacky}, {Labrie}, {Larkin}, {Maire}, {Marois}, {Millar-Blanchaer},
  {Nunez}, {Oppenheimer}, {Palmer}, {Pazder}, {Perrin}, {Poyneer}, {Quirez},
  {Rantakyro}, {Reshtov}, {Saddlemyer}, {Sadakuni}, {Savransky},
  {Sivaramakrishnan}, {Smith}, {Soummer}, {Thomas}, {Wallace}, {Weiss}, \&
  {Wiktorowicz}}]{Macintosh2012}
{Macintosh}, B.~A., {Anthony}, A., {Atwood}, J., {et~al.} 2012, in Proc. SPIE,
  Vol. 8446

\bibitem[{{Marois} {et~al.}(2006){Marois}, {Lafreni{\`e}re}, {Doyon},
  {Macintosh}, \& {Nadeau}}]{Marois2006}
{Marois}, C., {Lafreni{\`e}re}, D., {Doyon}, R., {Macintosh}, B., \& {Nadeau},
  D. 2006, \apj, 641, 556

\bibitem[{{Marois} {et~al.}(2008){Marois}, {Lafreni{\`e}re}, {Macintosh}, \&
  {Doyon}}]{Marois2008}
{Marois}, C., {Lafreni{\`e}re}, D., {Macintosh}, B., \& {Doyon}, R. 2008, \apj,
  673, 647

\bibitem[{{Marois} {et~al.}(2010){Marois}, {Macintosh}, \&
  {Veran}}]{Marois2010}
{Marois}, C., {Macintosh}, B., \& {Veran}, J.-P. 2010, Adaptive Optics Systems
  II. Edited by Ellerbroek, 7736, 52

\bibitem[{{Martinache} {et~al.}(2012){Martinache}, {Guyon}, {Clergeon},
  {Garrel}, \& {Blain}}]{Martinache2012}
{Martinache}, F., {Guyon}, O., {Clergeon}, C., {Garrel}, V., \& {Blain}, C.
  2012, in Proc. SPIE, Vol. 8447

\bibitem[{{Mawet} {et~al.}(2005){Mawet}, {Riaud}, {Absil}, \&
  {Surdej}}]{Mawet2005b}
{Mawet}, D., {Riaud}, P., {Absil}, O., \& {Surdej}, J. 2005, \apj, 633, 1191

\bibitem[{{Oppenheimer} {et~al.}(2012){Oppenheimer}, {Beichman}, {Brenner},
  {Burruss}, {Cady}, {Crepp}, {Hillenbrand}, {Hinkley}, {Ligon}, {Lockhart},
  {Parry}, {Pueyo}, {Rice}, {Roberts}, {Roberts}, {Shao}, {Sivaramakrishnan},
  {Soummer}, {Vasisht}, {Vescelus}, {Wallace}, {Zhai}, \&
  {Zimmerman}}]{Oppenheimer2012}
{Oppenheimer}, B.~R., {Beichman}, C., {Brenner}, D., {et~al.} 2012, in Proc.
  SPIE, Vol. 8447

\bibitem[{Quanz {et~al.}(2012)Quanz, Lafreni{\`e}re, Meyer, Reggiani, \&
  Buenzli}]{Quanz2012}
Quanz, S.~P., Lafreni{\`e}re, D., Meyer, M.~R., Reggiani, M.~M., \& Buenzli, E.
  2012, Astronomy and Astrophysics, 541, 133

\bibitem[{{Quanz} {et~al.}(2010){Quanz}, {Meyer}, {Kenworthy}, {Girard},
  {Kasper}, {Lagrange}, {Apai}, {Boccaletti}, {Bonnefoy}, {Chauvin}, {Hinz}, \&
  {Lenzen}}]{Quanz2010}
{Quanz}, S.~P., {Meyer}, M.~R., {Kenworthy}, M.~A., {et~al.} 2010, \apjl, 722,
  L49

\bibitem[{{Rousset} {et~al.}(2003){Rousset}, {Lacombe}, {Puget}, {Hubin},
  {Gendron}, {Fusco}, {Arsenault}, {Charton}, {Feautrier}, {Gigan}, {Kern},
  {Lagrange}, {Madec}, {Mouillet}, {Rabaud}, {Rabou}, {Stadler}, \&
  {Zins}}]{Rousset2003}
{Rousset}, G., {Lacombe}, F., {Puget}, P., {et~al.} 2003, in Proc. SPIE, ed.
  P.~L. {Wizinowich} \& D.~{Bonaccini}, Vol. 4839, 140--149

\bibitem[{{Soummer} {et~al.}(2012){Soummer}, {Pueyo}, \&
  {Larkin}}]{Soummer2012}
{Soummer}, R., {Pueyo}, L., \& {Larkin}, J. 2012, \apjl, 755, L28

\bibitem[{{Spiegel} \& {Burrows}(2012)}]{Spiegel2012}
{Spiegel}, D.~S. \& {Burrows}, A. 2012, \apj, 745, 174

\bibitem[{{Vigan} {et~al.}(2012){Vigan}, {Patience}, {Marois}, {Bonavita}, {De
  Rosa}, {Macintosh}, {Song}, {Doyon}, {Zuckerman}, {Lafreni{\`e}re}, \&
  {Barman}}]{Vigan2012}
{Vigan}, A., {Patience}, J., {Marois}, C., {et~al.} 2012, \aap, 544, A9

\end{thebibliography}

\end{document}